\documentstyle[epsfig]{elsart}
\begin{document}

\newcommand{\re}{\mathop{\mathrm{Re}}}
\newcommand{\im}{\mathop{\mathrm{Im}}}
\newcommand{\I}{\mathop{\mathrm{i}}}
\newcommand{\D}{\mathop{\mathrm{d}}}
\newcommand{\E}{\mathop{\mathrm{e}}}

\def\lambar{\lambda \hspace*{-5pt}{\rule [5pt]{4pt}{0.3pt}} \hspace*{1pt}}

{\Large  DESY 12-173}

{\Large  October 2012}

\bigskip

\bigskip

\bigskip

\begin{frontmatter}

\journal{Nucl. Instrum. and Methods A}

\date{}

\title{
A possible upgrade of FLASH for harmonic lasing down to 1.3 nm}

\author{E.A.~Schneidmiller}
\author{and M.V.~Yurkov}

\address{Deutsches Elektronen-Synchrotron (DESY),
Notkestrasse 85, D-22607 Hamburg, Germany}

\begin{abstract}
We propose the 3rd harmonic lasing in a new FLASH undulator as a way to produce
intense, narrow-band, and stable SASE radiation down to 1.3 nm with the present accelerator energy of 1.25 GeV.
To provide optimal conditions for harmonic lasing, we suggest to suppress the fundamental with the help of
a special set of phase shifters. We rely on the standard technology of gap-tunable planar hybrid undulators, and
choose the period of 2.3 cm and the minimum gap of 0.9 cm; total length of the undulator system
is 34.5 m. With the help of numerical simulations we demonstrate that
the 3rd harmonic lasing at 1.3 nm provides peak power at a gigawatt level and the narrow intrinsic bandwidth, 0.1\% (FWHM).
Pulse duration can be controlled in the range of a few tens of femtoseconds, and the peak brilliance reaches
the value of $10^{31}$ photons/($\mathrm{s \ mrad^2 \ mm^2 \ 0.1\% \ BW}$). With the given undulator design,
a standard option
of lasing at the fundamental wavelength to saturation is possible through the entire water window and at longer wavelengths.
In this paper we briefly consider additional options such as polarization control, bandwidth reduction,
self-seeding, X-ray pulse compression, and two-color operation.
We also discuss possible technical issues and backup solutions.
\end{abstract}

\end{frontmatter}

\baselineskip 20pt

\clearpage

\section{Introduction}

FLASH (Free electron LASer in Hamburg) is the first soft X-ray FEL user facility \cite{flash,njp} based
on Self-Amplified Spontaneous Emission (SASE) principle \cite{ks-sase}.
Presently, the facility is operated for users with the
accelerator energy up to 1.25 GeV and the shortest
wavelength of 4.1 nm \cite{siggi-fel}, i.e. the photon energy is slightly above Carbon K-edge.
This makes it possible to perform
first experiments in the so-called water window. However, even shorter wavelengths are requested by the FLASH user community.
In particular, lasing through the entire water window (i.e. down to 2.34 nm) would be interesting for some experiments.
Moreover,
resonant magnetic scattering studies would strongly profit from lasing down to 1.3 nm. In this case the L-edges
of the most interesting materials would be covered \cite{gr}. One of the possible ways to extend wavelength range
would be a major energy upgrade of FLASH \cite{ssy-fl,siggi-fl}. However,
such an extensive energy upgrade is not possible in the next few years. In this paper we propose an alternative way,
namely using the present accelerator energy of 1.25 GeV and the 3rd harmonic lasing in a new undulator.

Harmonic lasing in single-pass high-gain FELs \cite{murphy,hg-2,kim-1,mcneil,parisi,sy-harm}
is the radiative instability at an odd harmonic of the
planar undulator developing independently from lasing at the fundamental wavelength. Contrary to nonlinear harmonic
generation (which is driven by the fundamental in the vicinity of saturation),
harmonic lasing can provide much more intense, stable, and narrow-band FEL beam
which is easier to handle due to the suppressed fundamental. The most attractive feature of saturated harmonic lasing
is that the brilliance of a harmonic is comparable to (or even larger than) that of the fundamental. In
our recent study \cite{sy-harm} we came to the conclusion that the 3rd harmonic lasing in X-ray FELs is much more robust
than usually thought, and can be widely used at the
present level of accelerator and FEL technology. We surprisingly found out that in many cases the 3D model of harmonic
lasing gives more optimistic results than the 1D model. For instance, one of the results of our studies was that
in a part of the parameter space, corresponding to the operating range of soft X-ray beamlines of X-ray FEL facilities
(like SASE3 undulator of the European XFEL),
harmonics can grow faster than the fundamental mode. We also concluded that in many practical
cases the 3rd harmonic lasing has a
shorter saturation length than lasing at the same wavelength with the retuned fundamental mode (which is achieved by
opening the undulator gap or increasing electron energy).

For a successful harmonic lasing the fundamental mode must be suppressed.
A possible method to disrupt the fundamental without affecting the
third harmonic lasing was suggested in \cite{mcneil}: one can
use $2\pi/3$ phase shifters between undulator modules. We found out, however, that this method is
inefficient in the case of a SASE FEL (the simulations in \cite{mcneil} were done for the case of a monochromatic seed).
In \cite{sy-harm} we proposed a modification of the phase shifters method which can also work in the case of a SASE FEL.
We also proposed the suppression of the fundamental harmonic by using a spectral filter in a chicane installed
between two parts of the undulator. In this paper we consider an application of the modified phase shifter method for
the 3rd harmonic lasing in a new undulator at FLASH. An unusual feature of such an undulator is that one needs many
phase shifters (several tens) in order to disrupt the fundamental. Note that, in principle, one can also consider
harmonic lasing in FLASH II \cite{flash2} undulator.
However, a number of phase shifters is not sufficient there, and one would
need to replace one undulator module with a chicane and a filter. Also, neither the relatively long period of FLASH II
undulator, nor the fact that the electron beam must be deflected by a large angle (with a possible deterioration of beam
quality) support a possibility of lasing at short wavelengths under discussion.
Therefore, a better strategy would be to replace FLASH undulator after FLASH II is commissioned and started
user operation.

In this paper we show that the saturation of 3rd harmonic lasing at 1.3 nm can be achieved within 25 m (net magnetic
length) of the optimized undulator at FLASH if we assume that slice parameters of the electron beam are close to those
taken from start-to-end simulations \cite{s2e}. Saturated harmonic lasing would then provide a gigawatt power level,
narrow bandwidth (0.1\% FWHM), and a high peak brilliance,
exceeding $10^{31}$ photons/($\mathrm{s \ mrad^2 \ mm^2 \ 0.1\% \ BW})$. In the same undulator one can lase
to saturation at the fundamental wavelength through the entire water window.
In this paper we also briefly consider such
additional options as polarization control, bandwidth reduction, self-seeding, X-ray pulse compression, and two-color
operation. We discuss
possible technical issues, and come to the conclusion that all the necessary requirements to the accelerator system
and the undulator can be satisfied. We also consider frequency doubler as a backup solution for producing
soft X-ray radiation down to 1.3 nm.

\section{Parameters of the electron beam and the undulator}

A possibility of FLASH operation in the considered wavelength range (down to 1.3 nm) is supported by recent achievements
in production of low-emittance electron beams. This was demonstrated at LCLS \cite{lcls}, and also at PITZ test
facility \cite{pitz} which is responsible for the development of photoinjector technology for FLASH and the
European XFEL \cite{euro-xfel-tdr}. In particular, very low emittances (well below 1 $\mu$m)
were demonstrated for low charges \cite{pitz}.
Start-to-end simulations \cite{s2e} for FLASH and the European XFEL have shown that low emittances
can be preserved during bunch compression and transport of electron beams to the undulator. In Table 1 we present
parameters of the electron beam that were used in our FEL simulations (the results are shown below in this paper).
We used the model of Gaussian bunch with the charge
of 150 pC and slice parameters close to those obtained in start-to-end simulations \cite{s2e} in this range of charges.

\begin{table}[tb]
\caption{Parameters of electron beam and undulator}
\bigskip

\begin{tabular}{ l l }
\hline \\
{\bf Electron beam} & {\bf Value} \\
\hline \\
Energy & 1.25 GeV  \\
Charge & 150 pC \\
Peak current	 &  2.5 kA        \\
Rms normalized slice emittance	 &  0.5 $\mu$m     \\
Rms slice energy spread  &  250 keV   \\
Rms pulse duration & 24 fs \\
\hline \\
{\bf Undulator} & {\bf Value} \\
\hline \\
Period  & 2.3 cm \\
Minimum gap &  9 mm \\
$K_{\mathrm{rms}}$ (at minimum gap) &  1 \\
Beta-function & 7 m \\
Net magnetic length & 25 m \\
Total length & 34.5 m \\
\hline \\
\end{tabular}

\label{tab:param}
\end{table}

\begin{figure*}[t]

\includegraphics[width=1.\textwidth]{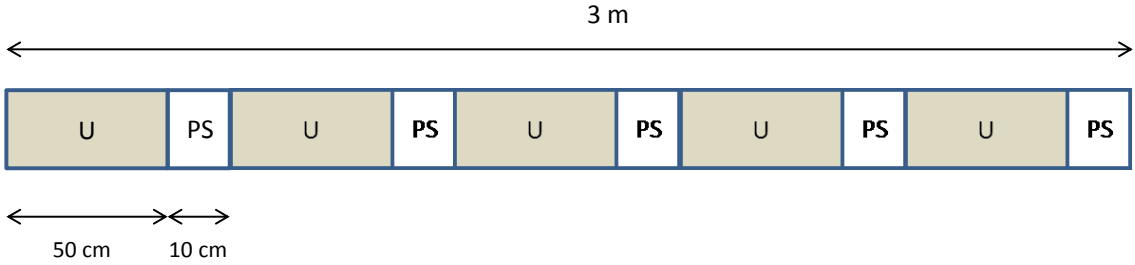}

\caption{\footnotesize Schematic view of an undulator module, consisting of undulator sections (U) and phase
shifters (PS).}

\label{module}
\end{figure*}

Parameters of the new undulator are also presented in Table 1. Several gap-tunable hybrid undulators with the period
of 2.3 cm are in operation at the Advanced Photon Source \cite{aps}, and for the gap of 9 mm the rms value of K
(which is equal to
peak value divided by $\sqrt{2}$) in the Table 1 is very safe and conservative \cite{gluskin}. As it was
already mentioned, an unusual feature of the proposed undulator is a big number of phase shifters that are necessary for
a sure suppression of the fundamental. We propose to build a 3 m long undulator module
consisting of 0.5 m long sections with
a phase shifter behind each section, see Fig.~\ref{module}. Phase shifters for this parameter range can be made
compact: either permanent magnet \cite{gluskin} or electromagnetic \cite{tischer} phase shifters can easily fit
in 10 cm space between
sections. An example of the compact design of the permanent magnet phase shifter for the Swiss FEL can be found in
\cite{swiss}, its length is less than 4 cm.
Note that the electromagnetic phase shifters are more preferable if one considers a fast switching (5 Hz) of
circular polarization (left-right) in cross-planar undulator (see discussion in Section 5).  We assume that the gap
between two modules, shown schematically in Fig.~\ref{module}, is 0.5 m long
(for quadrupoles, BPMs etc.), so that with 10 modules the total length of the
undulator system is 34.5 m, and the net magnetic length is 25 m.

\begin{figure*}[tb]

\includegraphics[width=.7\textwidth]{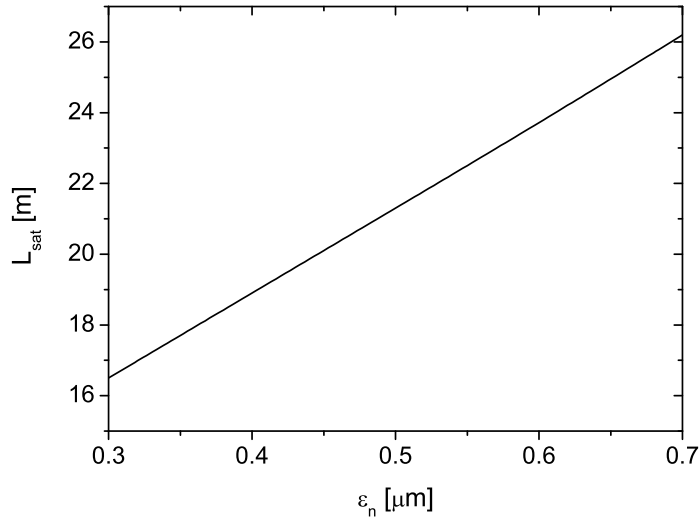}

\caption{\footnotesize Saturation length of the 3rd harmonic lasing at 1.3 nm versus normalized emittance of the electron beam.
Other parameters are given in Table 1.}

\label{sat-emit}
\end{figure*}

In Fig.~\ref{sat-emit} we present the dependence of the
saturation length of the 3rd harmonic lasing at 1.3 nm on the normalized emittance of the electron beam.
The calculations were done with the help of formulas presented in \cite{sy-harm}.
From this plot one can decide how much contingency might have to be added to the undulator length.

Closing this Section, we should explain why we need so many phase shifters.
If we want the third harmonic to lase to its saturation, we have to suppress the fundamental.
A method to disrupt the fundamental (while keeping the lasing at the third harmonic undisturbed) was
proposed in \cite{mcneil}. The undulators for X-ray FELs consist of many segments. In case of gap-tunable undulators, phase
shifters are foreseen between the segments. If phase shifters are tuned such that the phase delay is $2\pi/3$
(or $4\pi/3$) for the fundamental, then its amplification is disrupted. At the same time the phase shift is equal to $2\pi$
for the third
harmonic, i.e. it continues to get amplified without being affected by phase shifters.
However, the simulations in \cite{mcneil}
were done for the case of a monochromatic seed,
and the results cannot be applied for a SASE FEL. The reason is that
in the latter case the amplified frequencies are defined self-consistently, i.e. there is frequency shift (red or blue)
depending on positions and magnitudes of phase kicks. This leads to a significantly weaker suppression effect.
In particular, a consecutive
use of phase shifters with the same phase kicks $2\pi/3$  (as proposed in \cite{mcneil}) is inefficient \cite{sy-harm},
i.e. it does not lead to a sufficiently strong suppression of the fundamental wavelength. The method was generalized
in \cite{parisi},
where the alternation of the shifts $2\pi/3$ and $4\pi/3$ was considered. This improves the situation, but still does not
provide a sure suppression of the fundamental in realistic situations.

In \cite{sy-harm} we proposed another modification of phase shifters method that works better in the case of a SASE FEL.
We define phase shift in the same way as it was done in \cite{mcneil} in order to make our results compatible with the
previous studies.
For example, the shift $2\pi/3$ corresponds to the
advance\footnote{In a phase shifter (like a small magnetic chicane) the beam is, obviously, delayed
with respect to electromagnetic field. One can, however, always add or subtract $2\pi$, so that the shift is kept between 0 and
$2\pi$. Therefore, a delay in the phase shifter by $2\pi/3$ corresponds to the phase shift of $4\pi/3$ according to the
definition in \cite{mcneil}, and vice versa.}
of a modulated electron beam with respect to electromagnetic field by $\lambda/3$.
In the following we assume that a distance between phase shifters is shorter than the field gain length of the fundamental
harmonic.
Our method of disrupting the fundamental mode can be defined as a piecewise use of phase shifters with the strength
$2\pi/3$ and $4\pi/3$.
For example, in the first part of the undulator (consisting of several segments with phase shifters between them)
we introduce phase shifts $4\pi/3$.
A red-shifted (with respect to a nominal case without phase shifters) frequency band is amplified starting up from shot
noise. In the
following second part of the undulator we use $2\pi/3$ phase shifts, so that the frequency band, amplified in the first part,
is practically excluded from the amplification process. In a realistic 3D case,
the radiation is diffracted out of the electron beam, and
the density and energy modulations within this frequency band are partially suppressed due to emittance and energy spread
while the beam is passing the second part of the undulator (although the suppression effect is often not strong).
Instead, a blue-shifted frequency band is amplified in the second part of the undulator, starting up from shot noise.
Then, in the third part we change to no phase shifts case, then  again to $4\pi/3$ in the fourth part, etc.
A more complicated optimization can include using one or two $2\pi/3$ shifts in the first part with $4\pi/3$ phase shifters,
and so on. The efficiency of the method strongly
depends on the ratio of the distance between phase shifters and the field gain length of the undisturbed fundamental mode.
The smaller this ratio, the stronger
suppression can be achieved after optimization of phase shifts distribution. In Section 4 we present an example on the
optimized distribution of phase shifters. The result of our studies is that in the considered parameter range
we need a distance about 0.5 m between phase shifters.

\section{Wavelength tunability}

\begin{figure*}[tb]

\includegraphics[width=.7\textwidth]{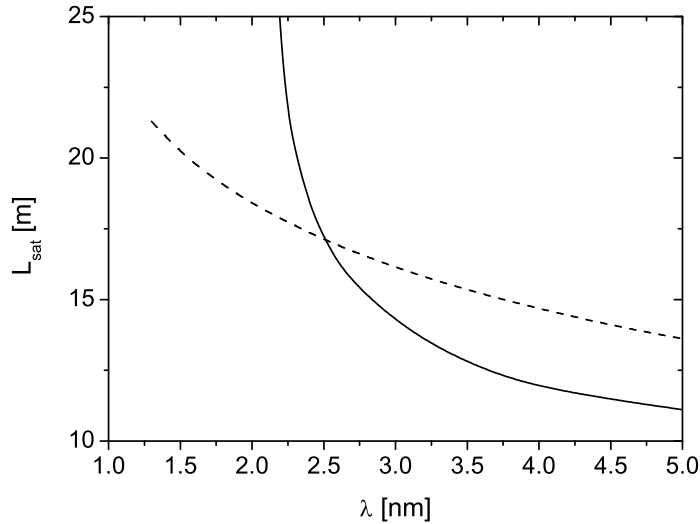}

\caption{\footnotesize Saturation length of the fundamental (solid) and of the 3rd harmonic (dash)  versus
wavelength. In the case of the 3rd harmonic the tuning of wavelength is done by changing beam energy for a fixed gap
with $K_{\mathrm{rms}}=1$. In the case of the fundamental the tuning is achieved by changing $K_{\mathrm{rms}}$
at fixed beam energy of 1.25 GeV for
wavelengths shorter than 3.9 nm, and by changing beam energy at $K_{\mathrm{rms}}=1$ for longer wavelengths.
In all cases the beam energy does not exceed 1.25 GeV. Other parameters are given in Table 1.}

\label{sat-wl}
\end{figure*}

Now we would like to consider a wavelength tunability for the undulator and beam parameters discussed in the previous
Section. The saturation length was calculated with formulas presented in \cite{sy-harm}. To obtain dependencies presented
In Fig.~\ref{sat-wl} we changed either electron energy (keeping it below 1.25 GeV) for $K_{\mathrm{rms}}=1$
(in the whole range for the 3rd harmonic and above 3.9 nm for the fundamental), or $K_{\mathrm{rms}}$ for the fixed energy
of 1.25 GeV (below 3.9 nm for the fundamental). In other words, we calculated the shortest saturation length for every
wavelength in the cases of lasing at the fundamental and at the 3rd harmonic.
It can be seen from Fig.~\ref{sat-wl} that one can reach saturation at the fundamental harmonic through
the entire water window, but for shorter wavelengths the saturation length quickly diverges because of too small value of
the undulator parameter $K_{\mathrm{rms}}$. Thus, shorter wavelengths can only be covered with the 3rd harmonic.
In the water window and at longer wavelengths the fundamental and the 3rd harmonic have comparable saturation lengths,
so that one can choose between two options. In the case of lasing at the fundamental
both the peak power and the bandwidth are larger by approximately a factor of 3-4. We did not show the
wavelengths longer than 5 nm in Fig.~\ref{sat-wl} but, obviously, lasing to saturation is easy to reach there.
We should also notice that the wavelength range can be slightly extended by 3rd harmonic lasing at 1.2 nm (by reducing the
undulator parameter), in this case the saturation length is still below 25 m. Thus, in terms of photon energy we can go
above 1 keV with the considered parameters of electron beam and undulator. In this paper we do not consider an energy
upgrade of FLASH, but it is worth noticing here that with the electron beam energy of 1.6-1.7 GeV one can lase to
saturation in the same undulator at 1.3 nm with the fundamental, and in sub-nanometer regime with the 3rd harmonic.

\section{Third harmonic lasing in a new undulator}

\begin{figure*}[tb]

\includegraphics[width=.7\textwidth]{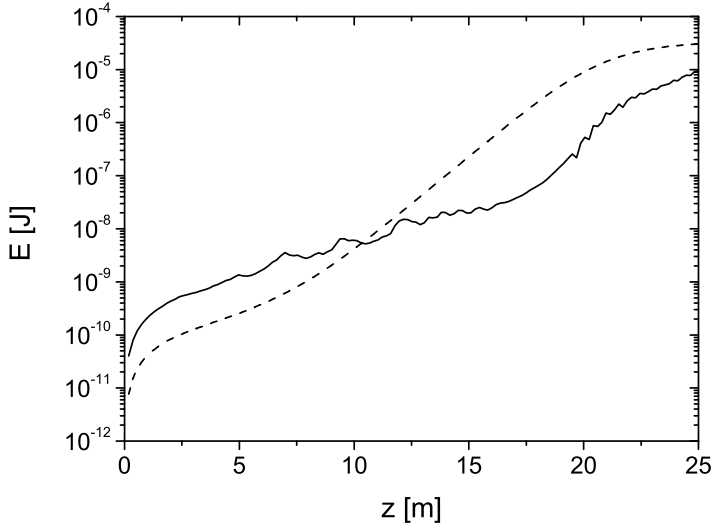}

\caption{\footnotesize Pulse energy versus magnetic length of the undulator for the fundamental (solid) and the 3rd harmonic
(dash). Electron beam and undulator parameters are given in Table 1. Phase shifters are located after every 0.5 m long
section of the undulator. The phase shift is $4\pi/3$ after sections 1-4, 6-9, 11-13, 18, 23, 39-49, and $2\pi/3$ after
sections 5, 10, 14-17, 19-22, 24-27.}

\label{gain-curve}
\end{figure*}

Third harmonic lasing to saturation at 1.3 nm is possible in the undulator described in Section 2. We performed
numerical simulations with the code FAST \cite{fast} recently adapted for harmonic lasing. In the simulations we used
the parameters presented in Table 1. In order to disrupt the fundamental harmonic with the help of phase shifters,
we used the approach described in Section 2. The phase shifters are positioned after every 0.5 m long section of
the undulator, and the distribution of the phase shifts is specified in the caption to Fig.~\ref{gain-curve}.
One can see from this Figure that the fundamental is sufficiently suppressed, and the 3rd harmonic can reach saturation
without being disturbed by the fundamental. Note that if the power of the fundamental (about 30\% of the
 3rd harmonic power in Fig.~\ref{gain-curve}) would hamper an experiment, it could
easily be filtered out by, for example, Aluminum filter.

\begin{figure*}[tb]

\centering
\includegraphics*[width=75mm]{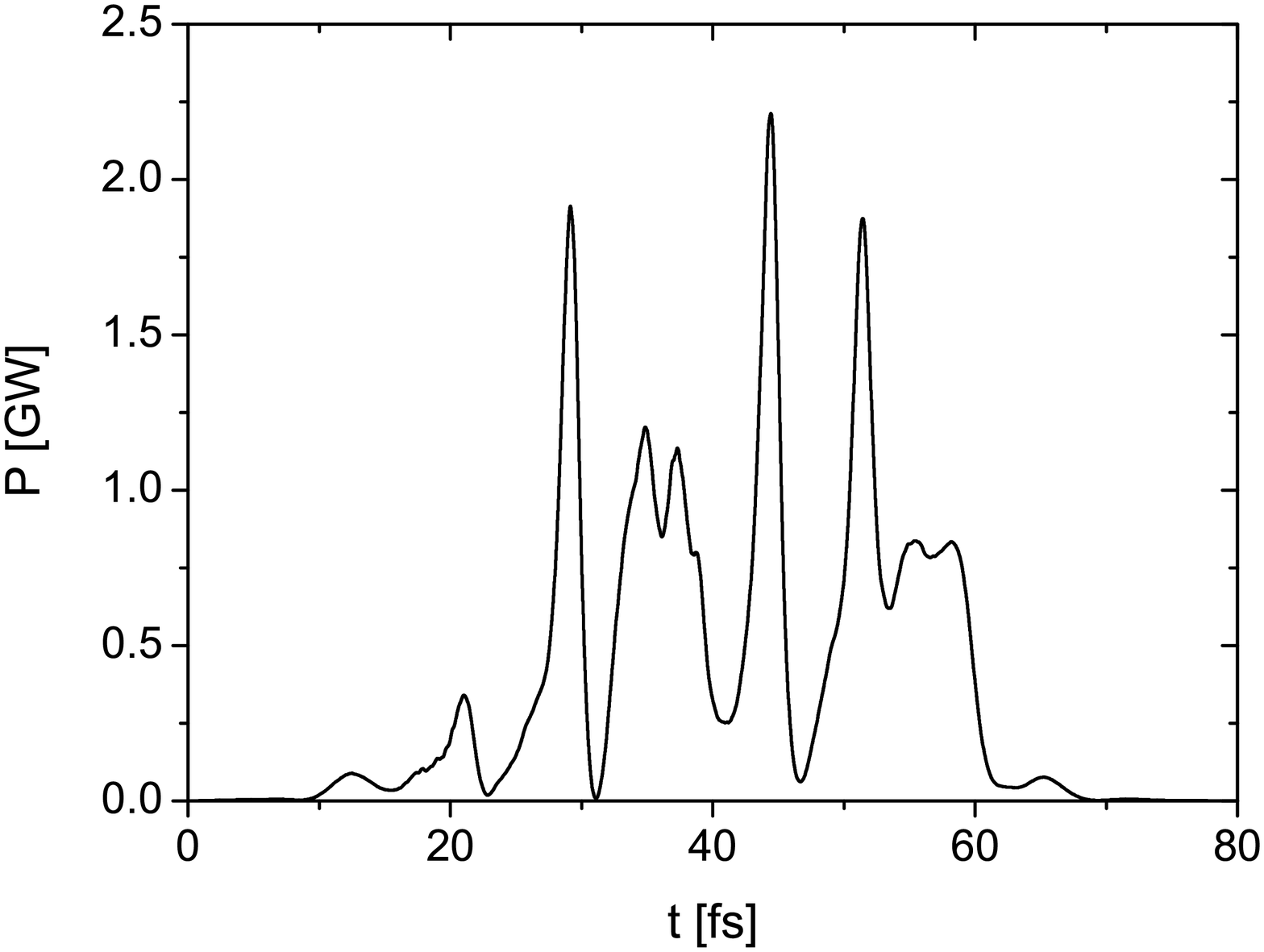}
\includegraphics*[width=75mm]{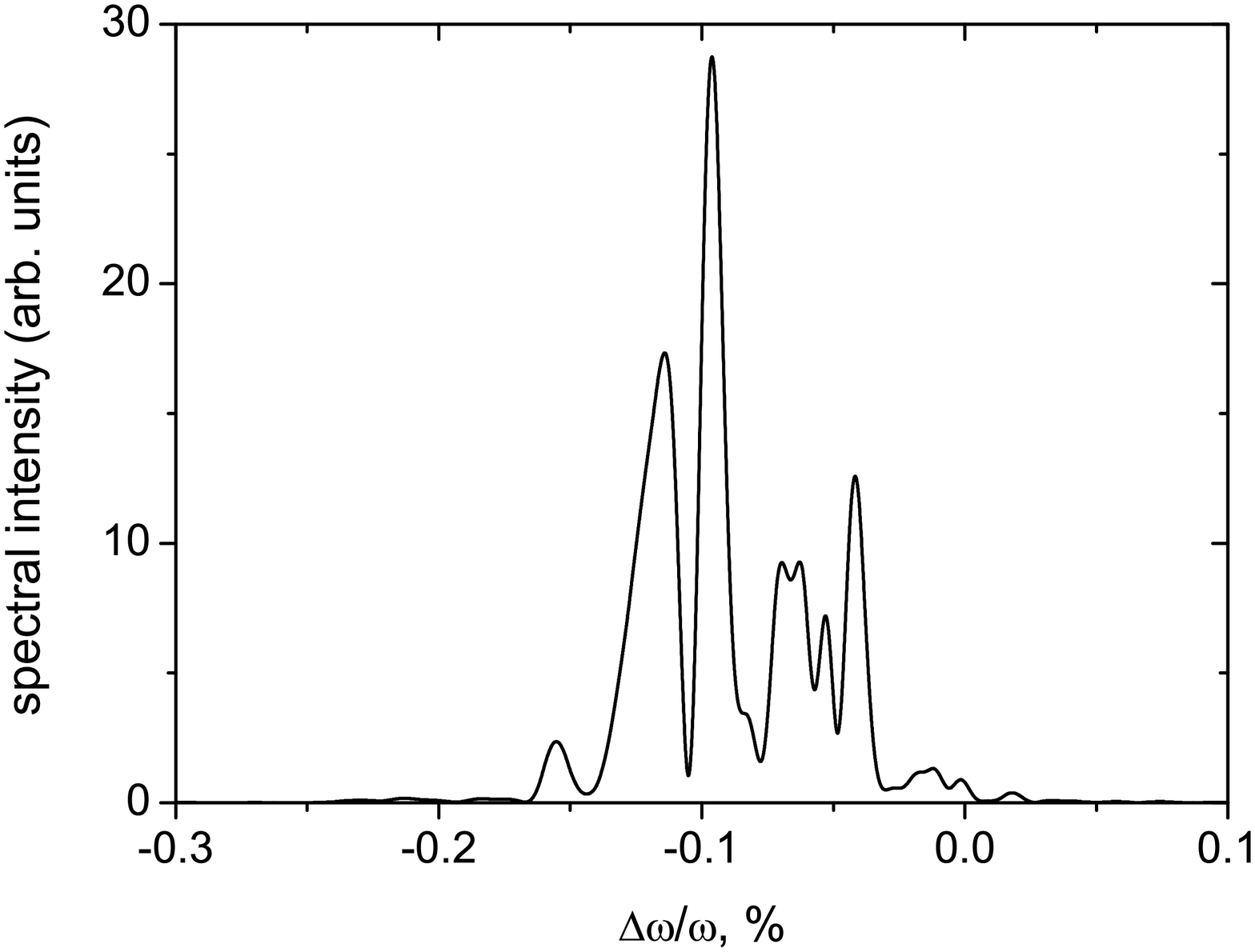}

\caption{\footnotesize
Single shot radiation power (left plot) of the 3rd harmonic at saturation and the corresponding spectrum (right plot).
Radiation wavelength is 1.3 nm. Electron beam and undulator parameters are given in Table 1.
}
\label{power}
\end{figure*}

In Fig.~\ref{power} we present radiation power of the third harmonic at saturation (single shot), and the
corresponding spectrum. One can see that the third harmonic lasing reaches gigawatt level of power, and also provides
a narrow bandwidth, about 0.1\% (FWHM).

Ensemble-averaged parameters of the radiation are summarized in Table 2. Note that pulse duration and pulse energy can
be varied within some range by changing, for example, bunch charge, while keeping slice parameters at the values close to
those from Table 1. Analyzing Table 2, we come to the conclusion that the third harmonic lasing to saturation
produces a very bright
photon beam. In particular, peak brilliance, a figure-of-merit of an X-ray FEL performance, may reach the record value for
FLASH. Parameters from Table 2 can satisfy many user experiments, in particular, resonant magnetic
scattering experiments \cite{gr}.

\begin{table}[b]
\caption{Parameters of radiation for saturated 3rd harmonic lasing}
\bigskip

\begin{tabular}{ l l }
\hline \\
Wavelength & 1.3 nm  \\
Averaged peak power & 1 GW \\
Pulse energy & 30 $\mu$J \\
Shot-to-shot fluctuations & $<$ 10 \% \\
Pulse duration (FWHM) & 30 fs \\
Bandwidth (FWHM) &  0.1 \% \\
Angular divergence (FWHM) &  10 $\mu$rad     \\
Peak brilliance  &  $10^{31} \ \mathrm{ph./ (s \ mrad^2 \ mm^2 \ 0.1\% \ BW}$)  \\
\hline \\
\end{tabular}

\label{tab:radiation}
\end{table}

\section{Additional options}

\subsection{Polarization control}

A circular polarization of the radiation is a preferred option for magnetic scattering experiments. Moreover,
one should be able to switch the polarization between left and right, especially attractive is a fast switching.
There are different approaches for production of circularly polarized radiation from X-ray FELs,
even in the case when the main undulator is planar. A possible option is the cross-planar undulator, originally proposed
for spontaneous radiation \cite{kim-pol-sr}, and later considered for a SASE FEL \cite{kim-pol-fel, huang-pol}.
The latter scheme
assumes using a short planar undulator, placed behind the main planar undulator, and having field polarization
perpendicular to that in the main undulator. A more advanced version of the cross-planar scheme was proposed for
spontaneous radiation \cite{tanaka}, and was recently suggested for the case of a SASE FEL \cite{geloni-pol}.
Instead of one short undulator,
the scheme uses several short undulators with alternating linear polarizations (vertical-horizontal) of magnetic field and
phase shifters between them.
Numerical simulations in  \cite{geloni-pol} demonstrate that in this case a relatively high degree of circular polarization
(above 95 \%) can be achieved at FEL saturation.

We notice here that such a scheme can, obviously, be generalized for operation with the third harmonic lasing in the planar
undulator. Moreover, the
proposed design of the undulator module supports naturally this option. If the last 3-m long undulator module
(see Fig. 1) is made of five 0.5 m long sections with alternating polarizations (H-V-H-V-H, assuming vertical polarization
in all other undulator modules), and the phase shifts
are optimized, the degree of circular polarization can be quite high (numerical simulations will be published elsewhere).
If we decide that the phase shifters should be electromagnetic, then 5 Hz switching between left and right circular
polarization would be possible.

\subsection{Bandwidth reduction}

As it was discussed above, harmonic lasing instead of lasing (at the same wavelength) at the fundamental may have an
advantage of a significantly narrower bandwidth which could be important for spectroscopic studies, for example.
As one can see from Fig.~\ref{sat-wl}, the gain lengths of the fundamental and of the third harmonic at the same wavelength
are comparable for the wavelengths above 2.5 nm. The bandwidth is smaller by a factor of 3-4 in the case of lasing at
the 3rd harmonic\footnote{For a  comparison of bandwidths one can use a simple relation, namely that the relative
bandwidth scales as $1/(L_{\mathrm{sat}} h)$, where $L_{\mathrm{sat}}$ is the saturation length and $h$ is a
harmonic number. Although this scaling is not universal, it holds in the considered range of parameters}. At longer
wavelengths the saturation length becomes shorter, and this leads  to an increase of the bandwidth.
In order to reduce bandwidth one can,
for example, increase beta-function or reduce peak current of the beam so that the saturation length becomes larger again.
Thus, the relative bandwidth at the level of
0.1 \% (FWHM) can be maintained in the large range of wavelengths. This would make it possible to use narrow-band
FEL radiation at all the beamlines, not only at PG2 (the beamline with plane grating monochromator).

\subsection{Self-seeding option}

For further reduction of the bandwidth a self-seeding scheme \cite{feldhaus} might be reconsidered\footnote{Unfortunately,
the
previous attempt \cite{treusch} to implement the self-seeding option was stopped despite the fact that the development was
at the advanced
stage.} for implementation at FLASH. Recent successful operation \cite{hxrss} at LCLS of the self-seeding scheme for hard
X-rays \cite{geloni}
demonstrates a potential of this approach for other facilities and wavelength ranges. Particularly
attractive for soft X-ray regime is the compact design \cite{sxrss} of the self-seeding setup proposed for LCLS.
A similar approach can be used at FLASH, in particular with the new undulator proposed in this paper. In this case one
undulator module is taken out, and is put upstream of the undulator (together with two or three additional modules).
It is substituted
by a chicane with the compact monochromator. Here we would like to mention that the self-seeding scheme will also work
in the case of harmonic lasing. Moreover, it is even more efficient since the total power loss through the
monochromator is reduced because of the narrower incoming bandwidth. In addition, the fundamental will be suppressed
more efficiently in the case of self-seeding (than in the case presented in Fig.~\ref{gain-curve}, for example)
because the monochromator will also serve as the filter for the fundamental \cite{sy-harm}.

\subsection{X-ray pulse compression}

A SASE FEL driven by the energy-chirped electron bunch produces a frequency-chirped radiation pulse. The latter can
be compressed with the help of dispersive elements (like a pair of gratings), also in the X-ray regime \cite{pell-comp}.
There is an idea to test this method of pulse length reduction at FLASH \cite{sasa}.
We note that the compression factor
depends on both the frequency chirp and the intrinsic FEL bandwidth. In other words, the bandwidth reduction by means
of harmonic lasing would help reduce the finite pulse length significantly (or reduce a required energy chirp for the
same compression factor), and a few femtosecond long pulses with the
enhanced peak power can be generated at FLASH. Note that if the energy chirp is strong enough to influence the gain
of harmonic lasing, this influence can be compensated for by the undulator taper \cite{chirp-tap}.

\subsection{Two-color operation}

The simultaneous lasing at the
fundamental wavelength and at the third harmonic with comparable intensities can be used in
jitter-free pump-probe experiments making use of a split-and-delay stage \cite{pp-harm}. For such an experiment
one can manipulate relative intensities with the help of phase shifters (see Fig.~\ref{gain-curve}).

One can also consider the simultaneous two-user operation.
In this case the electromagnetic phase shifters can have an advantage in comparison with permanent-magnet ones.
Indeed, the electromagnetic phase shifters can be switched from disrupting to non-disrupting state
between the macropulses: when they do not disrupt the fundamental,
it is delivered at the full power.
Thus, one can switch between two colors and then deliver them to different beamlines. The distribution between the
beamlines can be done with the help of movable mirrors \cite{tesla-rep} or dispersive elements.

\section{Possible issues and backup solutions}

One can identify possible technical issues \cite{sy-harm} using the fact that the harmonic lasing is more narrow-band
process than lasing at the fundamental wavelength (FEL parameter $\rho$ \cite{bon-rho} is smaller).
In other words, there are
more stringent requirements to the undulator field quality, stronger sensitivity to undulator wakefields etc.

In the considered numerical example of 3rd harmonic lasing at 1.3 nm, the effective $\rho$ is about $4 \times 10^{-4}$.
This value is unusually small for a soft X-ray FEL, but is typical for hard X-ray
machines \cite{lcls, euro-xfel-tdr, sacla, swiss}. In particular,
SACLA \cite{sacla} has a somewhat smaller value of  $\rho$, about $3 \times 10^{-4}$ at short wavelengths.
Such small values of
$\rho$ imply, in the first place, that the undulator must have a very good field quality.
Successful operation of the undulator
systems at LCLS and SACLA demonstrates that the required level of quality can be achieved for a hundred meter long
system, and even for a challenging technology of short-period in-vacuum undulators. In our proposal we rely on the standard
technology of out-of-vacuum hybrid planar undulators, and the undulator system is significantly shorter, so that
the required quality of the undulator is easier to achieve.

The wakefields in the undulator can have an impact on the amplification process in X-ray FELs \cite{stup},
a relevant parameter is a relative energy change per gain length, divided by $\rho$. However,
they can be compensated by using undulator
taper as it was successfully demonstrated at LCLS and at SACLA. The harmonic lasing is more sensitive to wakefields,
however a proper optimization of the taper allows to diminish this effect in many practical situations, especially for
low charge scenarios which we consider for 3rd harmonic lasing at FLASH.

We have shown in the paper that the application of harmonic lasing allows to generate FEL radiation with intrinsically
narrow bandwidth. However, from practical point of view it is important that the energy chirp in the electron beam and
energy fluctuations are smaller than a half of that bandwidth.
In the case of FLASH the energy chirps from RF and from the longitudinal space charge have opposite signs, and can to a
large extent compensate each other within the lasing part of the bunch \cite{s2e}. Minimization of the chirp can be done
empirically with the help of the transverse deflecting cavity, showing the longitudinal phase space of the electron beam
\footnote{On the other hand, in the case of X-ray pulse compression the required chirp can also be tuned.}.
As for the energy stability, shot-to-shot energy jitter at FLASH is rather small (especially for beam energies above 1 GeV),
on the order of $10^{-4}$ (rms), and long-term energy drifts can be compensated by the energy feedback and kept at the
required level.

The main uncertainty connected with harmonic lasing is that it has never been tested in high-gain FELs (while there were
several successful experiments on harmonic lasing in FEL oscillators \cite{benson-madey,warren,hajima,benson,sei,novofel}).
However, following our recent proposal \cite{sy-harm},
an experimental test at LCLS of the harmonic lasing in hard X-ray regime
is under discussion \cite{ratner}.

As a backup solution for reaching 1.3 nm in the proposed undulator, one can consider nonlinear generation of the
second harmonic. Indeed, if the first part of the undulator is tuned for lasing at 2.6 nm at the fundamental, the second
harmonic bunching (at 1.3 nm) occurs as soon as FEL process enters nonlinear regime. At that position one
increases the undulator parameter such that the fundamental is in resonance with 3.9 nm. Then the second harmonic of
density bunching produces radiation at the third harmonic of the undulator (1.3 nm). The radiation power is larger than
in a simple case of nonlinear generation of the 3rd harmonic in the undulator with constant undulator parameter because the
second harmonic of the density bunching is significantly higher than the third one.

As a more efficient option one can consider a frequency doubler \cite{doubler} in which a dispersive section is used to
improve the performance. The advantage of using the dispersive section is that one gets a strong density bunching
at the second harmonic with a
low level of energy modulations in the first part of the undulator. Then, in the second part of the undulator, tuned to
the second harmonic, the density modulation exists longer and produces more power because of small energy modulations.
As an option one can consider undulator taper in the second part of the undulator \cite{doubler}.
Note that the dispersive section
can be realized as a closed bump over two undulator modules with the help of moving quadrupoles.
Since the required value of
the $R_{56}$ is a few micrometers, it is sufficient to displace the electron beam by a few millimeters (the two undulators
can be used to generate additional $R_{56}$, it can be as large as 1-2 micrometers). We can make a simple estimate for
the lengths of different sections in this scheme: six undulator modules (net magnetic length is 15 m)
for lasing at 2.6 nm, two modules for the dispersion section, and two modules (5 m) for the
harmonic radiator at 1.3 nm (with the fundamental wavelength at 3.9 nm).
The option of producing circularly polarized light (as discussed above) also exists in the
case of frequency doubler.

\section{Conclusion}

In this paper we have considered a new option for reaching high photon energies (about 1 keV) at FLASH with the
present accelerator energy of 1.25 GeV. This is the 3rd harmonic lasing  which allows to obtain
photon beam with ultimate brilliance. We have discussed different additional options as well as possible technical issues
and backup solutions. Our conclusion is that it is worth seriously considering the 3rd harmonic lasing in a new
undulator at FLASH in the near future.

\section{Acknowledgments}

We would like to thank R. Brinkmann, W. Decking, B. Faatz, J. Feldhaus,
E. Gluskin, G. Gr\"ubel, T. Limberg, S. Schreiber, M. Tischer,
M. Vogt, and E. Weckert  for useful discussions.

\end{document}